\documentclass{PoS}
\usepackage{epsfig}
\usepackage{amsmath}
\usepackage{bm}
\usepackage{times}
\newcommand{\ens}{\epsilon_{ns}}
\newcommand{\es}{\epsilon_{s}}

\newcommand{\veff}{\mathcal{V}_{\mathit{eff}}}
\newcommand{\aeff}{\mathcal{A}_{\mathit{eff}}}
\newcommand{\vfus}{\mathcal{V}^{us}_\mathit{eff}}
\newcommand{\vfud}{\mathcal{V}^{ud}_\mathit{eff}}
\newcommand{\afus}{\mathcal{A}^{us}_\mathit{eff}}
\newcommand{\afud}{\mathcal{A}^{ud}_\mathit{eff}}

\newcommand{\gf}{1-\xi^2 \rho_L}

\def\roughly#1{\mathrel{\raise.3ex\hbox{$#1$\kern-.75em%
\lower1ex\hbox{$\sim$}}}}

\title{Did one observe couplings of right-handed quarks to W ?}

\ShortTitle{Right-handed currents}

\author{\speaker{Jan STERN}\\
        Institut de Physique Nucl\'eaire, Orsay, France\\
        E-mail: \email{stern@ipno.in2p3.fr}}

\abstract{I consider non standard EW couplings of light right-handed
quarks to W and Z suggested in a systematic non decoupling bottom-up low-energy
effective theory approach to possible extensions of the Standard Model. New experimental tests in $K^L_{\mu3}$ decays
based on recent measurements and scalar form factor analysis are discussed. A
successful NLO fit to the standard set
of Z-pole and other NC data is presented as well.}

\FullConference{KAON International Conference\\
		 May 21-25 2007\\
		 Laboratori Nazionali di Frascati dell'INFN, Rome, Italy}

\begin{document}

\section{Introduction}

        In the Standard Model (SM), right handed fermions do not couple to W
and their couplings to Z are proportional to the electric charge. Compelling
tests of this feature exist for leptons, whereas for quarks available tests
are less conclusive due to the interference with non perturbative QCD effects. 
Another characteristics of the right-handed sector of the SM is a rather
complicated and apriori unexplained spectrum of weak hypercharges. (Since
the seventies, the latter has motivated left-right symmetric extensions 
of the SM~\cite{LR} which shed a new light on
the EW couplings of right handed fermions.) None of the
above features of the SM follow from the EW symmetry
$S_{EW}= SU(2)_W \times U(1)_Y$, as long as the latter is spontaneously broken:
Indeed, with the help of agents of Symmetry Breaking (Higgs fields), it is  possible    
to construct $S_{EW}$ invariant couplings of right handed
fermions to W. This fact suggests to look for eventual  modifications of the  
right-handed couplings as a 
conceivable signal of a non standard EW symmetry breaking.
Model independent tests of EWSB require first of all a 
``bottom-up'' Effective Theory approach which starts from the known vertices
of the SM and step
by step in a low-energy expansion controlled by a {\bf power counting}
orders possible non standard effects according to their importance at low
energies. Next, it should be specified how the lepton - quark universality
could be naturally broken at subleading orders to escape  strong
experimental constraints concerning leptons. 

                            Such a class of LEETs has been proposed three
years ago~\cite{HS1} and further developed and completed later~\cite{HS2}. 
In this talk, I am
going to review the characteristic feature of this class: The appearance at
NLO of couplings of right
handed quarks to W and modification of their couplings to Z. Then I will
comment on first attempts to confront these predictions with
experiment~\cite{BOPS}.

\section{Not quite Decoupling EW Low Energy Effective Theory  (LEET)}

In its minimal version, the LEET contains the naturally light particles of the
SM: SU(2) x U(1) gauge fields, chiral fermions (including right-handed
neutrinos) and the triplet of GBs. For small 
momenta $p \ll 4\pi F_W = \Lambda_W \sim 3 TeV$, the effective Lagrangian is 
written as a low-energy expansion

\begin{equation}
\mathcal{L}_{\mathit{eff}} = \sum_{d \ge 2} \mathcal{L}_{d} , \quad 
\mathcal{L}_{d}  =  \mathcal{O}([p/\Lambda_W]^d)~,
\end{equation}

\noindent where the infrared dimension  of a local operator, 
$d = n_{\delta} + n_g  +  n_f/2$ is 
given by the number of derivatives, the number of gauge couplings and the 
number of fermion fields.  A Feynman diagram  with effective vertices v=1...
and with L loops counts at low-energy as O($p^d$), where
\begin{equation}
d = 2 + 2 L + \sum_{v} (d_v - 2)~.
\label{powercounting}
\end{equation}

\noindent The LEET is renormalizable order by order in the LE expansion, provided at 
each order, all terms allowed by symmetries are effectively included in (2.1). 
In particular,  the symmetry of the LEET $S_{nat} \supset S_{EW}$ must prevent
all ``unwanted `` non standard vertices to appear already at the leading order
$\mathcal{O}(p^2)$. In a bottom-up approach, the higher symmetry $S_{nat}$
is unknown apriori (it is the remnant of the not quite decoupled 
high energy sector of the theory), but it can be inferred requiring that
the leading order $\mathcal{L}_{2}$ of the LEET coincides with the
Higgs-free  part of the SM
Lagrangian. I refer to~\cite{HS2}, where it is shown that the {\bf minimal solution}
of this condition reads
\begin{equation}
S_{\mathit{nat}} = \left [ SU(2)_{G_L} \times SU(2)_{G_R} \times U(1)^{B-L}_{G_B} \right ]_{\mathit{elem}}
\times \left [ SU(2)_{\Gamma_L} \times SU(2)_{\Gamma_R} \right ]_{\mathit{comp}}~.
\label{snat}
\end{equation}
The Goldstone boson matrix $\Sigma(x) \in SU(2)$ (needed to give masses to W
and Z) transforms according to a different local chiral symmetry
\begin{equation}
\Sigma(x) \to \Gamma_L(x) \Sigma(x) \left[\Gamma_R(x)\right ]^{-1}
\end{equation}
than the chiral fermion doublets and the elementary gauge fields coupled to
fermions
\begin{equation}
 \psi_{L/R} \to G_{L/R}\exp\left[-i \frac{B-L}{2} \alpha \right] \psi_{L/R}~.
\label{trafofermions}
\end{equation}
The most general Lagrangian of dimension $d=2$ invariant under the linear
action of the symmetry $S_{nat}$ reads
\begin{eqnarray}
  \mathcal{L} \left( p^2 \right)  &=& \frac{F_W^2}{4}  \left\langle D_{\mu}
  \Sigma^{\dag} D^{\mu} \Sigma \right\rangle + i\, \overline{\psi_L}
  \gamma^{\mu} D_{\mu} \psi_L + i\, \overline{\psi_R} \gamma^{\mu} D_{\mu}
  \psi_R \nonumber \\
  && -  \frac{1}{2}  \left\langle G_{L \mu \nu} G_L^{\mu \nu} + G_{R \mu \nu}
  G^{\mu \nu}_R \right\rangle - \frac{1}{4} G_{B \mu \nu} G_B^{\mu \nu}~.  
  \label{lag}
\end{eqnarray}
It contains several gauge fields not observed at low energies $E< \Lambda_{W}$,
no fermion masses and a gauge boson mass term which has no obvious connection
with the SM. Nevertheless, the above Lagrangian reduces to the one of the SM
upon imposing $S_{nat}$ - invariant constraints eliminating the redundant
gauge fields through pairwise identification of different gauge factors
up to a gauge transformation. (Notice that these constraints break the
accidental L-R symmetry present in (2.6).) Example of such constraints is

\begin{equation}
\Gamma_{L,\mu} =  \mathcal{X} g_L G_{L,\mu} \mathcal{X}^{-1} +
i  \mathcal{X}  \partial_{\mu}  \mathcal{X}^{-1}
\label{xconstraint}
\end{equation}
which replaces $SU(2)_{G_{L}} \times SU(2)_{\Gamma_{L}}$ by its diagonal
subgroup (identified with the SM weak isospin) and a scalar object 
$\mathcal{X}$ which is a (constant) multiple of a $SU(2)$ matrix, and is
called ``spurion''. Similarly, one identifies up to a gauge
$\Gamma_{R,\mu} \sim g_{R} G_{R,\mu} \sim g_{B} G_{B,\mu} \tau_{3}/2$.
We then remain with the gauge fields of the SM, receiving standard masses
and mixing through the first term in (2.6) and coupled in the standard way
to fermions. In addition, we now have three $SU(2)$ valued spurions           
$\mathcal{X}$,  $\mathcal{Y}$ and $\omega$
\begin{equation}
\mathcal{X}(x) = \xi \Omega_L(x),\quad     \Omega_L(x) \in SU(2), \quad 
\mathcal{Y} = \eta\, \Omega_R,\quad \Omega_R \in SU(2), \quad 
\omega = \zeta \,\Omega_{B},\quad    \Omega_{B} \in SU(2)~, 
\end{equation}
populating the coset space $S_{nat}/S_{EW} = SU(2)^3$. To maintain
invariance under  $S_{nat}$, the spurions have to transform as
\begin{equation}
\mathcal{X} \to \Gamma_{L} \mathcal{X} G^{-1}_{L}, \quad 
\mathcal{Y}  \to \Gamma_R \, \mathcal{Y}\,  G_R^{-1}, \quad 
\omega  \to \Gamma_{R}\, \omega \, G^{-1}_{B} .
\end{equation}
Consequently, the constraints selecting $S_{EW} = SU(2)_W \times U(1)_{Y}$ of
the SM as the maximal subgroup of $S_{nat}$ that is linearly realized at low 
energies can be equivalently written as

\begin{equation}
D_{\mu}  \mathcal{X} = 0 ,\quad 
D_{\mu} \mathcal{Y}  =  0 ,\quad
D_{\mu} \omega       = 0 ~
\end{equation}
indicating that spurions do not propagate. There exists a gauge in which
the spurions reduce to three real parameters $\xi$, $\eta$ and $\zeta$
which are exterior to the SM and whose magnitude is not fixed by the LEET.
They will be considered as {\bf small expansion parameters} describing
effects beyond the SM.

The physical origin of spurions satisfying the constraints (2.10)
 can be understood as resulting from a particular non decoupling limit
 of an ordinary Higgs mechanism in which both Higgs bosons and some
combinations of gauge fields become very massive. Massive gauge fields
decouple, whereas heavy Higgs fields reduce to non propagating spurions, defining a 
non linear realization of the symmetry $S_{nat}/S_{EW}$. 

               Spurions are {\bf needed} to write down $S_{nat}$ invariant
 fermion masses.  Consequently, the latter will be suppressed with respect
to the scale $\Lambda_{W}$ by powers of spurion parameters $\xi$ and $\eta$.
The least suppressed mass - the top mass -  will be proportional to the
product            
\begin{equation}
\xi \eta \sim m_{top} / \Lambda_{W} = \mathcal{O}(p) , \quad
d^*  =  d  + \frac{1}{2} ( n_{\xi}  +  n_{\eta} ).
\end{equation}
This suggests to extend the low-energy power counting to spurions introducing
the chiral dimension $d^{\star}$ defined above. This guarantees that both
the fermion mass term and the lagrangian (2.6) have $d^{\star} = 2$
characteristic of the leading order of the LEET. Notice that the power
counting formula also holds replacing in (2.2) $d$ by $d^{\star}$.         
          
	  The third spurion $\omega$ 
breaks B-L, which is thus predicted to be a part of the LEET. Consequently,
the parameter $\zeta\ll\xi \sim \eta$ naturally accommodates the smallness
of Lepton number violation and of the Majorana masses.

\section{Next to  Leading Order  (NLO)}

            The NLO consists of all $S_{nat}$ invariant operators of the
 chiral dimension $d^{\star} = 3 $ . There are two and only two such
operators: they describe non standard couplings of fermions to W and Z
and they are suppressed by two powers of spurions $\mathcal{X}$ or
$\mathcal{Y}$ :
\begin{equation}
\mathcal{O}_L = \bar \psi_L \mathcal{X}^{\dagger}\gamma^\mu \Sigma D_{\mu}\Sigma^{\dagger}\mathcal{X}\psi_L ,
\end{equation}
for left handed fermions, whereas for right handed fermions one has
\begin{equation}
\mathcal{O}_R^{a,b} = \bar\psi_R \mathcal{Y}^{\dagger}_{a} \gamma^\mu \Sigma^{\dagger}
D_{\mu}\Sigma \mathcal{Y}_{b}\psi_R .
\end{equation}
where $a,b \in [U ,D]$, label  covariant projections on Up and 
Down components of right handed doublets.  These operators already carry their
respective suppression factors, they are $\mathcal{O}(p^2 \xi^2)$ and
$\mathcal{O}(p^2 \eta^2)$ respectively. The full $d^{\star} = 3$ part of
the effective Lagrangian can be written as
\begin{equation}
\mathcal{L}_{\text{NLO}} = \rho_L \mathcal{O}_L(l) + \lambda_L \mathcal{O}_L(q)+
\sum_{a,b}  \rho_R^{a,b} \mathcal{O}^{a,b}_R(l) + \sum_{a,b}  
\lambda_R^{a,b} \mathcal{O}_R^{a,b}(q)
\end{equation}
where $\rho$ and $\lambda$ are dimensionless low-energy constants which
should be of order one (unless suppressed by an additional symmetry).
The NLO couplings still respect the family symmetry . On the other hand,
at this subleading order, the lepton - quark universality could be broken,   
i.e. $\rho \neq \lambda$
by
the existence of additional reflection symmetry  $\nu_{R} \to -\nu_{R}$
which does not exist for quarks. Such a symmetry is not obstructed by the LO
couplings to gauge fields (at LO, $\nu_{R}$ decouples). It allows the right
handed neutrino to get a {\bf small Majorana mass} of the order          
$\mathcal{O}(\zeta^2 \eta^2)$, i.e. of a comparable size to left handed
Majorana mass $\mathcal{O}(\zeta^2 \xi^2)$ and to the strength of LNV.         
On the other hand, the reflection symmetry $\nu_{R} \to - \nu_{R}$ forbids
the Dirac neutrino-mass and could provide a natural explanation of the observed
smallness of neutrino masses. A corollary of this ``anti see-saw''
mechanism~\cite{HS2} of suppression of neutrino masses is the        
{\bf suppression of charged leptonic right-handed currents}
i.e. $\rho_{R}^{UD} = 0$ in Eq (3.3).

\section{Couplings to W}

                     Let us concentrate on couplings of fermions to
W. Using the matrix notation in the family space
$\mathrm{U}= (u,c,t)^T,~\mathrm{D}= (d,s,b)^T,~\mathrm{N}= (\nu_e, \nu_\mu, 
\nu_\tau)^T,~\mathrm{L}= (e, \mu, \tau)^T $ 
and using the mass - diagonal basis,  the couplings
   to W up to and including  NLO become
\begin{eqnarray}
\mathcal{L}_{\text{W}} &=& \frac{e (\gf)}{\sqrt{2} s}   
\left\{ \bar{\mathrm{N}}_L V_{\mathrm{MNS}} \gamma^{\mu} L_L 
+(1+\delta) 
\bar{\mathrm{U}}_L V_L \gamma^{\mu} D_L + 
\epsilon
\bar{\mathrm{U}}_R V_R \gamma^{\mu} D_R\right\}
W_{\mu}^+ + \text{h.c}~.\nonumber \\
\end{eqnarray}
$V_{L}$ and $V_{R}$ are two independent {\bf unitary}  mixing
matrices resulting (as in SM) from the diagonalization of quark masses.
The (small) spurionic parameters $\delta = (\rho_L - \lambda_L) \xi^2$ and
$\epsilon = \lambda_R ^{UD} \eta^2$ describe the chiral generalization of
the CKM mixing induced by RHCs. Notice, in particular, that effective
EW couplings in the vector and axial channels (more directly accessible
than $V_L$ and $V_R$)
\begin{eqnarray}
\veff^{ij}&=&(1+\delta) V_L^{ij}+\epsilon V_R^{ij}+\mathrm{NNLO}~,\nonumber\\
\aeff^{ij}&=&-(1+\delta) V_L^{ij}+\epsilon V_R^{ij}+\mathrm{NNLO}~,
\end{eqnarray}
need not to be unitary. The signal of RHCs can be detected as
$\veff^{ij} \neq - \aeff^{ij}$, i.e. comparing vector and axial vector
transitions.

                  A particular attention should be payed to {\bf light
 quarks $u, d, s$} for which the chirality breaking effects are tiny.
 In this sector all EW effective couplings can be expressed in terms of 
 $\delta$ and three parameters
\begin{equation}
\ens= \epsilon\ \mathrm{Re}
\Bigl{(}\frac{V_R^{ud}}{V_L^{ud}}\Bigr{)}, \quad
\es =\epsilon\ 
\mathrm{Re} \Bigl{(}\frac{V_R^{us}}{V_L^{us}}\Bigr{)}~, \quad
\vfud = 0.97377(26)\equiv cos\hat{\theta}
\end{equation}
where $\veff^{ud}$ is determined from $0^+ \to 0^+$ nuclear 
transitions~\cite{PDG}.
Using further the unitarity of $V_{L}$ and neglecting $|V_{L}^{ub}|^2$,
all light quark effective couplings can be expressed as
\begin{eqnarray}
|\vfud|^2 &=& \cos^2\hat\theta \nonumber\\
|\afud|^2 &=& \cos^2\hat\theta\,( 1 - 4 \, \ens) \nonumber\\
|\vfus|^2 &=& \sin^2\hat\theta\, (1 +
 2\frac{\delta+\ens}{\sin^2\hat\theta}) (1 + 2\, \es - 2\, \ens) \nonumber\\
|\afus|^2 &=& \sin^2\hat\theta\, (1 +
 2\frac{\delta+\ens}{\sin^2\hat\theta}) (1 -2\, \es - 2\, \ens)~.
\end{eqnarray}
 
                     The genuine spurion parameters $\delta$ and $\epsilon$
are expected to be at most of order few percent. Since             
$|V_L^{us}| \ll |V_{L}^{ud}| \sim 1 $ and the matrix $V_{R}$ is unitary,
one should have $|\epsilon_{NS}|< \epsilon$. On the other hand, the parameter
$\epsilon_{S}$ measuring RHCs strangeness changing transitions {\bf can
be enhanced} if the mixing hierarchy for right handed light quarks is inverted, 
$V_R^{ud} < V_{R}^{us}$. In this case $|\epsilon_{S}|$ could be as large
as $4.5 \epsilon$. Clearly, this question should be decided experimentally.

 \section{The stringent test of RHCs: Scalar $K_{\mu3}$ form factor shape}

                              Model independent bounds on $V+A$ couplings of
  light quarks to W are extremely difficult to find, since they require an
  accurate control of QCD chiral symmetry breaking contributions when
comparing
  hadronic matrix elements of vector and axial vector currents. One such test
  (never considered before) has been identified in Ref~\cite{bops}. It is based
  on the Callan Treiman low - energy Theorem already discussed in the talk by
E. Passemar~\cite{emilie}. The (normalized) scalar $K^L_{\mu3}$ form factor 
$f(t)$
                  
\begin{equation}
f(t)=\frac{f^{K^0\pi^-}_S(t)}{f^{K^0\pi^-}_+(0)} = \frac{1}{f^{K^0\pi^-}_+(0)}
\left(f^{K^0\pi^-}_+(t) + \frac{t}{\Delta_{K\pi} } f^{K^0\pi^-}_-(t)\right)
\ \ ,\ \ f(0)= 1 .
\end{equation}
where $\Delta_{K\pi} = m^2_{K} - m^2_{\pi}$, satisfies

\begin{equation}
C \equiv f(\Delta_{K\pi})= \frac{F_{K^+}}{F_{\pi^+}}\frac{1}{f_{+}^{K^0\pi^-}
(0)}+  \Delta_{CT}, \quad
C = B_{\mathit{exp}} \, r + \Delta_{CT}.
\end{equation}
Here, $\Delta_{CT} = - 3.5 \times 10 ^ {-3}$ is a tiny correction which has
been estimated in one loop ChPT.
In the absence of RHCs, the value C of the scalar form factor at the Callan
Treiman point can be directly expressed in terms of measured branching
fractions ($K_{l2}/\pi_{l2}$, $K_{l3}$) and $V^{ud}$ giving~\cite{PDG}
$B_{exp}=1.2438 \pm 0.0040$ in the second Eq.~(5.2). RHCs make appear 
additional correction factor $r$

\begin{equation}
r = \Bigl{|}\frac{\afud\vfus}{\vfud \afus}\Bigr{|} = 1 + 2 (\epsilon_{S} -
\epsilon_{NS} ).
\end{equation} 
 Hence,  in the presence of RHCs the Callan Treiman theorem yields
\begin{equation}
\ln C  = 0.2182 \pm 0.0035 + \tilde\Delta_{CT} + 2 (\es -
\ens) = 0.2182 \pm 0.0035 + \Delta\epsilon~,
\end{equation}
with $\tilde\Delta_{CT} = \Delta_{CT} / B_{exp}$.

                                                         An accurate
 physically motivated parametrization of the scalar form-factor $f(t)$
 has been proposed~\cite{bops} which allows to determine the parameter $\ln C$ from the measured
 $K^L_{\mu3}$ decay distributions. The corresponding measurement is
particularly delicate, since the experimental t - distribution is not easy to
reconstruct from the data. Furthermore, different experiments have access to different
decay distributions which do not have the same sensitivity to $\ln C$ and to
the shape of the vector form factor. There exists a relation between
$\ln C$ and the slope parameter $\lambda_0$ ~\cite{emilie} but it is not enough
precise to reduce the determination of $\ln C$ to existing (controversial)
determinations of the slope $\lambda_0$ assuming the linear t-dependence of
the scalar form factor~\cite{KLOE, KTeV, Lai} or at most injecting information about its 
curvature~\cite{LeptonPhotonKLOE}. Recently, NA48 collaboration has published
the result of a direct determination of $\ln C$ based on the dispersive
representation of $f(t)$ ~\cite{Lai}
\begin{equation}
\ln C_{\mathit{exp}} = 0.1438 \pm 0.0138, \quad
\Delta\epsilon = -0.074\pm0.014
\end{equation}
Other analysis of $K_{\mu3}$ decay distributions from KLOE~\cite{Antonelli} and KTeV~\cite{Glazov} based on
the dispersive representation of the two form factors are underways. They should clarify the experimental situation and
provide an independent cross check of the NA48 result~\cite{Lai}. 
             Awaiting an independent dispersive analysis of existing data samples,
 one should stress that the result (5.5) indicates a 5$\sigma$ deviation
from the SM prediction. In particular, if the discrepancy would have to be
explained within QCD, the ChPT estimate of $\Delta_{CT}$ would have to be
underestimated by a factor 20. On the other hand, within the class of LEET 
defined above the interpretation of the result (5.5) as a manifestation of 
couplings of right handed quarks to W is unambiguous. It amounts to a 
determination of the spurion parameter $2 (\epsilon_{S} - \epsilon_{NS})$. 
Its size can be understood as a result of
enhancement of $V_{R}^{us}$ relative to the suppressed $V_{L}^{us}$.
Beyond our LEET framework, other interpretations might
be conceivable. For example a subTeV charged scalar coupled to scalar
densities $\bar{u}s$ and $\bar{\mu}\nu$ could interfere with our analysis.
We prefer to stay within the class of minimal LEET defined above and ask
how does the same non standard operator (3.2) affect the couplings of right
handed quarks to Z.
                                  
\section{Couplings to $Z$}
                     
		     Non standard couplings to Z contained in the NLO
Lagrangian (3.3) are suppressed by the same two spurion parameters $\xi^2$
(LHCs) and $\eta^2$ (RHCs) as in the case of couplings to W discussed in
Section 4. Hence, despite the apriori unknown ``order one'' prefactors
$\rho$ and $\lambda$, it is possible to relate orders of magnitude of
non standard CC and NC couplings. In the left - handed sector we have
altogether two NLO parameters: $ \delta = \xi^2 (\rho_L - \lambda_L)$
and $ \xi^2 \rho_L $, whereas in the right - handed sector there are three
new parameters denoted $\epsilon^{e},~\epsilon^{U},~\epsilon^{D}$
and proportional to the spurion $\eta^2$.

\begin{figure}[h]
\begin{center}
\includegraphics*[scale=0.35]{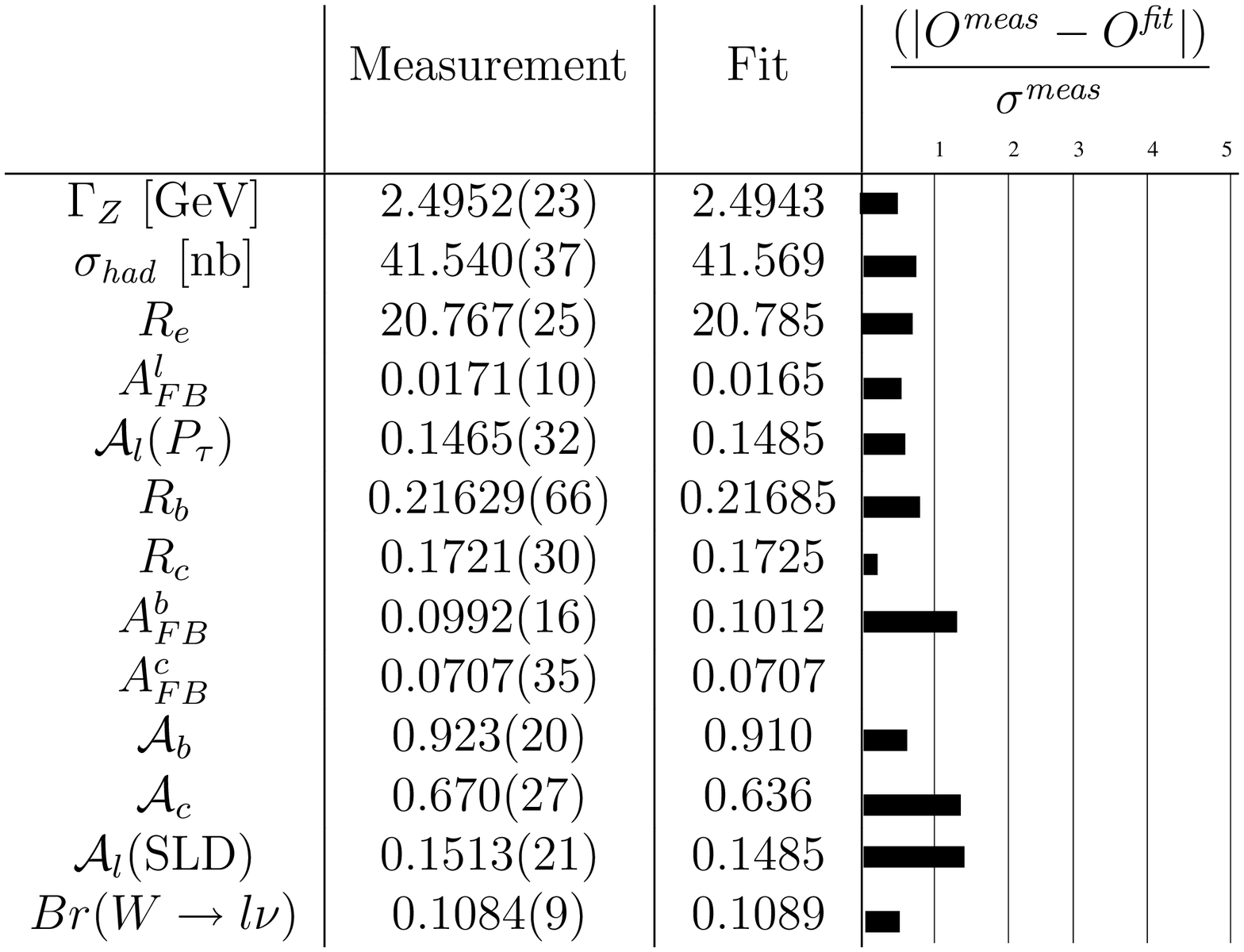}
\caption{\it Pull for the $Z$ pole observables in the full fit }
\end{center}
\end{figure} 

                         We have performed the NLO fit to the usual set of Z
- pole pseudo-observables displayed in Fig. 1 including the lepton branching
fraction of W (particularly sensitive to the parameter $\delta$) as well as
spin asymmetries measured at SLD.  The fit is described in details in
\cite{BOPS}. It has $\chi^2 /dof = 8.5/8 $ and it gives
$ \delta \equiv \xi^2 (\rho_{L} - \lambda_{L}) = - 0.004(2)$,~
$\xi^2 \rho_{L} = 0.001(12)$ and
$\epsilon^{e} \equiv \eta^2 \rho_{R}^{DD} = -0. 0024(5)$.\\
The most important NLO modification of couplings to Z turns out to
occur for right handed quarks:
$\epsilon^{U} \equiv \eta^2 \lambda_R^{UU} = -0.02(1)$ ~~
$\epsilon^{D} \equiv \eta^2 \lambda_R^{DD} = - 0.03(1)$.
The correlations can be found in \cite{BOPS}.

                         Two comments are in order. First, the most
important NLO non standard couplings to Z seem to occur for {\bf right
handed quarks}. Their size compares well with the couplings of right handed
quarks to W as suggested by the $K^L_{\mu3}$ dispersive Dalitz plot analysis
\cite{Lai}. Next, the fit is of a very good quality as illustrated in
Fig. 1 in terms of ``pulls''. In particular, the b-quark forward backward
asymmetry $A^b_{FB}$ {\bf and} $R^{b}$ are both well reproduced 
without modifying the flavour universality of NC EW couplings. The long
standing ``puzzle of b-asymmetries'' has apparently gone thanks to
the modified right-handed couplings of D type quarks to Z.

\section{$F_K/F_{\pi}$ and $f_{+}(0)$}
The low-energy QCD quantities $F_K$, $F_{\pi}$, $f_+(0)$ \ldots 
are defined independently of EW interactions in terms
of QCD correlation functions and they are accessible to
ChPT and lattice studies. On the other hand their experimental values
extracted from semileptonic branching fractions depend on the presumed EW
vertices via the effective EW couplings (4.4). 
Fixing experimental values
of $\vfud$ (4.3) and of the semi leptonic branching ratios,
$F_K , F_{\pi}, f_+(0)$ \ldots become unique functions of spurion parameters
$\epsilon_{NS}$, $\epsilon_S$ and $\delta$. One has
\begin{equation}
\left(\frac{F_{K^+}}{F_{\pi^+}}\right)^2 
=\left (\frac{\hat{F}_{K^+}}{\hat{F}_{\pi^+}}
\right)^2 \frac{1+2\,(\es-\ens)}{1+\frac{2}{\sin^2\hat{\theta}}
(\delta+\ens)},\ \
|f^{K^0\pi^-}_+(0)|^2 = \left[ \hat{f}^{K^0\pi^-}_+(0)\right]^2 \,
\frac{1-2(\es-\ens)}{1+\frac{2}{\sin^2\hat{\theta}}
(\delta+\ens)},
\end{equation}

\noindent where the hat indicates the corresponding values extracted from semi
leptonic branching fractions ({\bf {assuming SM couplings
$\epsilon_{NS}=\epsilon_S=\delta=0$}}). The latter are known very precisely:

\begin{eqnarray}
\hat{F}_{K^+}/\hat{F}_{\pi^+} = 1.182(7), ~~ 
\hat{f}^{K^0\pi^-}_{+} (0) = 0.951(5)~.
\end{eqnarray}

\begin{figure}[t]
\begin{center}
\epsfig{width=10cm,file=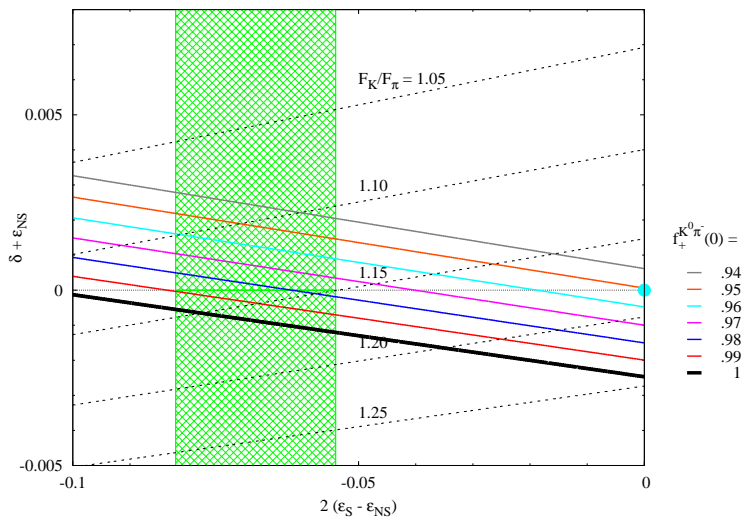}
\caption{\it {Lines of constant values for $F_{K+}/F_{\pi+}$ and
$f_+^{K^0\pi}(0)$ in the plane $\delta+\ens$ and $2 (\es-\ens)$ as resulting from Eqs (7.1) and (7.2). 
The vertical band indicates the
range suggested by the NA48 measurement~\cite{Lai}. The SM point $\epsilon=\delta =0$ is
also shown.}}
\end{center}
\end{figure}
\noindent In fig. 2 are displayed lines of constant values of $F_K/F_{\pi}$ 
and $f_+(0)$ as a function of spurion parameters. One notes
that $F_K/F_{\pi}$ significantly decreases compared with the value
1.22 often used as input in ChPT. On the other hand, $f_+(0)$
is not very constrained despite the Callan Treiman relation.
Finally, nothing prevents the effective vector mixing matrix $\mathcal{V}_{eff}$ to be
nearly unitary without any fine tuning. One has
\begin{equation}
| \vfud|^2 + |\vfus|^2
= 1 + 2 \, (\delta + \ens) +             
2\, (\es-\ens)\, \sin^2\hat{\theta}~.
\label{effunit}
\end{equation}
The contribution of $\epsilon_S$, the only parameter which
might be enhanced above 0.01 is suppressed by $sin^2\hat{\theta}$.
\\

      In conclusion, we have presented and motivated a
new low-energy test of non standard EW couplings of
right handed quarks not considered before. Did one observe
couplings of right handed quarks to W in $K^L_{\mu3}$ decay ? The final answer
requires a more complete and dedicated experimental analysis. It also
deserves a particular effort despite its difficulty. 
\\   
   
       I thank V. Bernard, M. Oertel, and E. Passemar for a valuable
collaboration. This work has been supported in part by the EU contract MRTN-CT-2006-035482 (Flavianet).

\end{document}